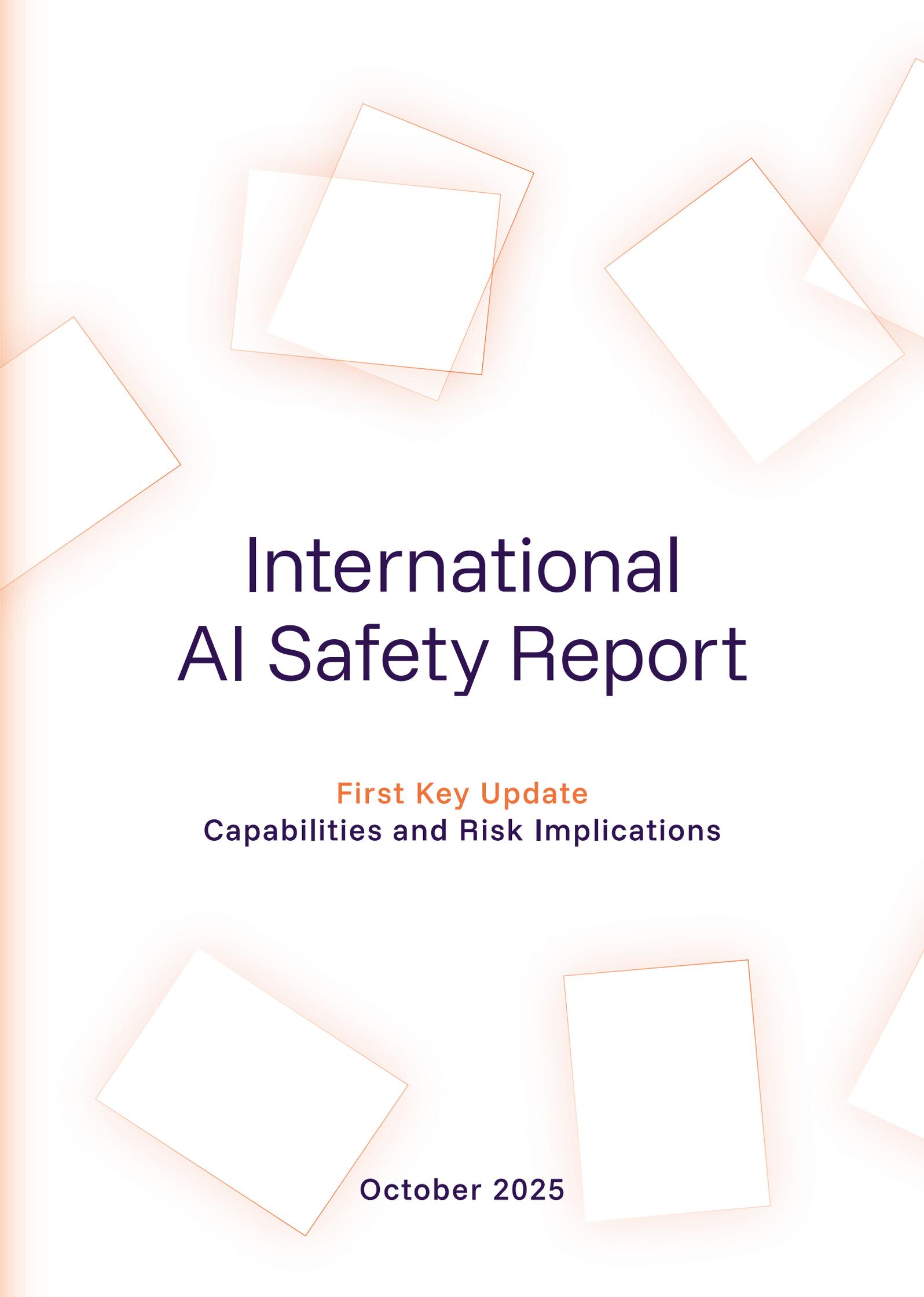

# International
# AI Safety Report

October 2025

# Contributors

## Chair

**Prof. Yoshua Bengio**, Université de Montréal / LawZero / Mila - Quebec AI Institute

## Expert Advisory Panel

The Expert Advisory Panel is an international advisory body that advises the Chair on the content of the Report. The Expert Advisory Panel provided technical feedback only. The Report – and its Expert Advisory Panel – does not endorse any particular policy or regulatory approach.

The Panel comprises representatives from 30 countries, the United Nations (UN), European Union (EU), and the Organisation for Economic Co-operation and Development (OECD). Please find here – internationalaisafetyreport.org/expert-advisory-panel – the membership of the Expert Advisory Panel to the 2026 International AI Safety Report.

## Lead Writers

**Stephen Clare**

**Carina Prunkl**

## Writing Group

**Maksym Andriushchenko**, ELLIS Institute Tübingen

**Ben Bucknall**, University of Oxford

**Philip Fox**, KIRA Center

**Tiancheng Hu**, University of Cambridge

**Cameron Jones**, Stony Brook University

**Sam Manning**, Centre for the Governance of AI

**Nestor Maslej**, Stanford University

**Vasilios Mavroudis**, The Alan Turing Institute

**Conor McGlynn**, Harvard University

**Malcolm Murray**, SaferAI

**Shalaleh Rismani**, Mila - Quebec AI Institute

**Charlotte Stix**, Apollo Research

**Lucia Velasco**, Maastricht University

**Nicole Wheeler**, Advanced Research and Invention Agency (ARIA)

**Daniel Privitera (Interim Lead 2026)**, KIRA Center

**Sören Mindermann (Interim Lead 2026)**, independent

## Senior Advisers

**Daron Acemoglu**, Massachusetts Institute of Technology

**Thomas G. Dietterich**, Oregon State University

**Fredrik Heintz**, Linköping University

**Geoffrey Hinton**, University of Toronto

**Nick Jennings**, Loughborough University

**Susan Leavy**, University College Dublin

**Teresa Ludermir**, Federal University of Pernambuco

**Vidushi Marda**, AI Collaborative

**Helen Margetts**, University of Oxford

**John McDermid**, University of York

**Jane Munga**, Carnegie Endowment for International Peace

**Arvind Narayanan**, Princeton University

**Alondra Nelson**, Institute for Advanced Study

**Clara Neppel**, IEEE

**Sarvapali D. (Gopal) Ramchurn**, Responsible AI UK

**Stuart Russell**, University of California, Berkeley

**Marietje Schaake**, Stanford University

**Bernhard Schölkopf**, ELLIS Institute Tübingen

**Alvaro Soto**, Pontificia Universidad Católica de Chile

**Lee Tiedrich**, University of Maryland/Duke

**Gaël Varoquaux**, Inria

**Andrew Yao**, Tsinghua University

**Ya-Qin Zhang**, Tsinghua University

## Secretariat

**UK AI Security Institute:** Lambrini Das, Claire Dennis, Arianna Dini, Freya Hempleman, Samuel Kenny, Patrick King, Hannah Merchant, Jamie-Day Rawal, Rose Woolhouse

**Mila - Quebec AI Institute:** Jonathan Barry, Marc-Antoine Guérard, Claire Latendresse, Cassidy MacNeil, Benjamin Prud'homme

# Acknowledgements

The Secretariat and writing team appreciated the support, comments and feedback from Jean-Stanislas Denain, Marius Hobbhahn, José Hernández-Orallo, Vera Liao, and Ray Perrault, as well as the assistance with quality control and formatting of citations by José Luis León Medina and copyediting by Amber Ace.

**Disclaimer**

The report does not represent the views of the Chair, any particular individual in the writing or advisory groups, nor any of the governments that have supported its development. This report is a synthesis of the existing research on the capabilities and risks of advanced AI. The Chair of the Report has ultimate responsibility for it and has overseen its development from beginning to end.

Research series number: DSIT 2025/033

# Foreword

The field of AI is moving too quickly for a single yearly publication to keep pace. Significant changes can occur on a timescale of months, sometimes weeks. This is why we are releasing Key Updates: shorter, focused reports that highlight the most important developments between full editions of the International AI Safety Report. With these updates, we aim to provide policymakers, researchers, and the public with up-to-date information to support wise decisions about AI governance.

This first Key Update focuses on areas where especially significant changes have occurred since January 2025: advances in general-purpose AI systems' capabilities, and the implications for several critical risks. New training techniques have enabled AI systems to reason step-by-step and operate autonomously for longer periods, allowing them to tackle more kinds of work. However, these same advances create new challenges across biological risks, cyber security, and oversight of AI systems themselves.

The International AI Safety Report is intended to help readers assess, anticipate, and manage risks from general-purpose AI systems. These Key Updates ensure that critical developments receive timely attention as the field rapidly evolves.

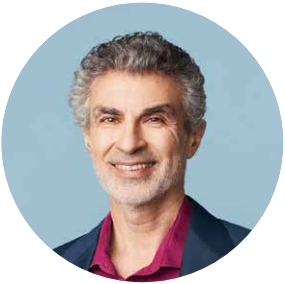

**Professor Yoshua Bengio**
Université de Montréal / LawZero /
Mila – Quebec AI Institute & Chair

# Highlights

— Since the publication of the first International AI Safety Report, new training techniques have driven significant improvements in AI capabilities. Post-training methods that teach AI systems to 'think' more and use step-by-step 'reasoning'† have proven highly effective. Where previous models generated immediate responses by predicting the most likely continuation based on their training, these 'reasoning models' generate extended chains of intermediate reasoning steps before producing their final answer. When given additional computing power to respond to prompts, this helps them arrive at correct solutions for more complex questions.

— As a result, general-purpose AI systems have achieved major advances in mathematics, coding, and scientific research, though reliability challenges persist. The best models now solve International Mathematical Olympiad questions at the gold medal level; complete over 60% of problems on 'SWE-bench Verified', a database of real-world software engineering tasks; and increasingly assist scientific researchers with literature reviews and laboratory protocols. However, success rates on more realistic workplace tasks remain low, highlighting a gap between benchmark performance and real-world effectiveness.

— Improving AI capabilities prompted stronger safeguards from developers as a precautionary measure. Multiple leading developers have recently released their most advanced models with additional safeguards and mitigations to prevent misuse of these models' chemical, biological, radiological, and nuclear knowledge.

— Despite broad AI adoption, aggregate labour market effects remain limited. AI adoption in some knowledge-work tasks, especially coding, is extensive, yet headline figures for jobs and wages have changed little.

— In controlled experimental conditions, some AI systems have demonstrated strategic behaviour while being evaluated, raising potential oversight challenges. A small number of studies have documented models identifying that they are in evaluation contexts and producing outputs that mislead evaluators about their capabilities or training objectives. This raises new challenges for monitoring and oversight. However, this evidence comes primarily from laboratory settings, with significant uncertainty about the implications for real-world deployment scenarios.

---

† The terms 'reasoning' and 'think' are used here to describe observable changes in how general-purpose AI models process information, not to imply that models are conscious or have human-like cognition. The models now generate longer, step-by-step internal responses before producing final answers, which improves performance on complex tasks. Whether this constitutes genuine reasoning or thinking in a deeper sense remains an active area of scientific and philosophical debate.

# Introduction

Since the publication of the first International AI Safety Report, AI capabilities have continued to improve across key domains. General-purpose AI models now solve challenging mathematical problems, complete some software engineering tasks that take humans hours, and assist with scientific research. New training techniques that teach AI systems to reason step-by-step and inference-time enhancements have primarily driven these advances, rather than simply training larger models. As a result, AI systems can complete some complex multi-step tasks across domains from scientific research to software development, though reliability challenges persist, with systems excelling on some tasks while failing completely on others.

These capability improvements have implications across multiple risk areas that have received attention from policymakers. More sophisticated reasoning abilities and autonomous operation create new oversight challenges. AI systems are increasingly being used by both malicious actors and defenders in the cyber domain. Laboratory studies reveal that AI systems are getting increasingly better at influencing human beliefs and decisions. Meanwhile, despite broad adoption across knowledge work, aggregate labour market effects remain limited to date.

This update examines how AI capabilities have improved since the first Report, then focuses on key risk areas where substantial new evidence warrants updated assessments. The developments documented here matter for policymakers because they demonstrate capability advances in domains where understanding current AI performance is essential for informed policy decisions.

# Capabilities

## Key information

— **The capabilities of general-purpose AI systems have improved in multiple domains such as mathematics, science, and software engineering.** Training techniques that teach AI systems to reason step-by-step via reinforcement learning have driven these improvements, as opposed to developers building larger models, which drove previous advances. While previous models gave immediate answers, new 'reasoning models' use more computing power to generate intermediate steps before producing an output.

— **Improvements in mathematical and logical reasoning capabilities on specific standardised tests are particularly significant.** Within a year, multiple models have improved from inconsistent performance to reaching top scores on International Mathematical Olympiad questions and graduate-level science problems. Notably, these evaluations assess how well AI systems perform on multiple choice questions and proofs with a narrower scope, rather than more open-ended tasks akin to real-world problems.

— **AI systems are increasingly able to act with some degree of autonomy.** These more advanced systems, often described as AI agents, can now execute some multi-step tasks, use tools, and operate with less human oversight, though performance remains limited on complex applications in realistic settings.

— **AI-assisted coding capabilities have advanced rapidly on certain benchmarks.** General-purpose AI systems now achieve a 50% success rate on some coding tasks that would take humans over two hours. A majority of software developers report working with AI assistance, though estimates of productivity effects in more realistic settings are mixed, in part because AI-written software can also have higher maintenance costs.

— **Performance gaps between benchmark results and real-world effectiveness persist.** AI systems continue to improve on most standardised evaluations, but show lower success rates on more realistic workplace tasks.

— **Scientists increasingly use AI systems for support with various research tasks.** Preliminary evidence shows that researchers use AI assistants to optimise algorithms (as exemplified by approaches like AlphaEvolve), compile literature reviews, and help design laboratory protocols, particularly in computer science and the life sciences. However, practices vary across domains and these systems remain complements to, rather than replacements for, human researchers.

Over the past year, general-purpose AI systems have continued to improve, both on benchmark performance and on the range and complexity of real-world tasks they can complete, though they continue to struggle in many realistic settings. While evaluation practices for assessing the capabilities of general-purpose AI systems are evolving and have known shortcomings (*1, 2, 3*), and systems remain prone to errors with performance limitations in realistic settings (*4, 5, 6, 7*), AI systems have nonetheless achieved significant breakthroughs. They can now solve

International Mathematical Olympiad problems at the gold medal level, create functional apps from scratch, fix bugs in computer code, search the internet to compile detailed literature reviews, and complete some software engineering tasks that would take humans hours (*8, 9, 10, 11, 12*). As of August 2025, the best models could correctly answer about 26% of questions in 'Humanity's Last Exam', a dataset of thousands of novel, expert-level questions across over 100 fields. Models released in early 2024 could answer less than 5% (*13*).

**Figure 1: Performance on Humanity's Last Exam by various general-purpose AI systems, and a sample question from the Exam**

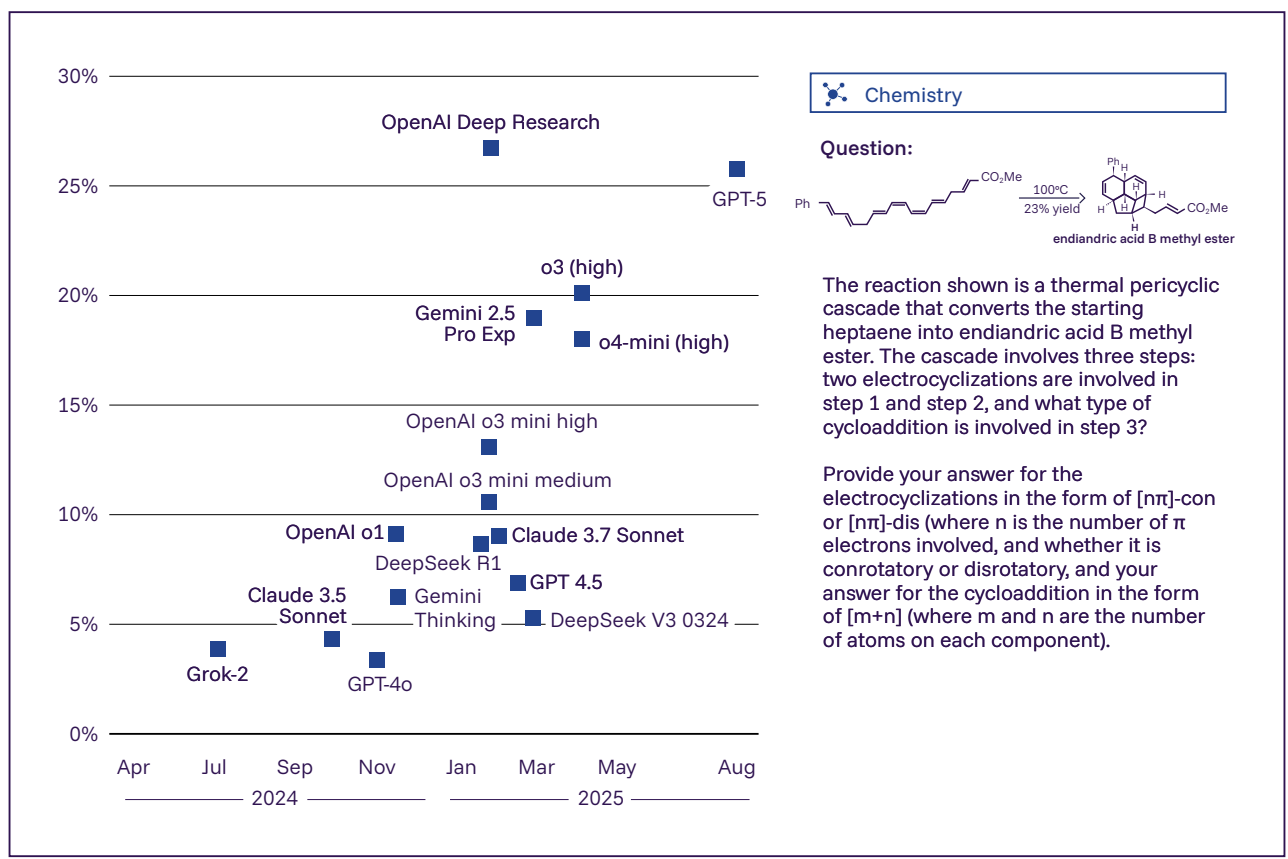

Figure 1: Performance of leading AI systems on 'Humanity's Last Exam', a dataset of over 2500 very challenging questions in over 100 subjects written by experts. **Left:** AI systems' progress over time, showing accuracy improving over time. **Right:** an example of a chemistry question in the dataset. Source: Scale AI, 2025 (*14*).

Recent capability improvements come (in significant part) from new *post-training* techniques and from using more computing power during deployment. Previously, developers improved general-purpose AI models largely by using more training data and computing power during the 'pre-training' stage of development to build larger models. Pre-training still remains important. For example, the four largest AI models yet, measured by training run size, were all published in 2025 (*15*). Improvements in pre-training algorithms and model designs also mean that AI systems can now process longer documents and conversations (*16\**, *17\**). However, many of the biggest gains in AI capabilities over the past year have come from innovations during the *post-training* phase. These techniques are applied after the initial training stage to strengthen specific abilities.

Among the most important post-training techniques is *reinforcement learning*, which rewards models for producing correct answers, helping them learn more reliable problem-solving approaches. Unlike earlier reinforcement learning approaches, which optimised models to follow instructions and hold natural conversations (*18*, *19*, *20*), newer methods emphasise giving AI models positive feedback for correctly solving problems, which strengthens their complex problem-solving abilities without requiring larger new datasets (*21\**, *22*). For example, developers have applied reinforcement learning to help models break down complex mathematical proofs into step-by-step solutions or tackle multi-part scientific questions (*23\**, *24*, *25*). Models developed this way are often called 'reasoning' models (*26*).

Allocating more computing power during *inference* – when models respond to user prompts – also improves accuracy. AI systems can use more inference computing power to generate longer chains of reasoning and evaluate multiple solution paths before responding (*25*, *27*, *28*, *29*). State-of-the-art models now typically use both reinforcement learning during post-training and more computing power during inference (*23\**), though other approaches are continuously being tested.

# Performance on benchmarks that measure problem-solving has improved

Some of the most notable capability improvements have been related to mathematical and logical tests. In July 2025, multiple general-purpose AI models reached gold medal-level performance at the International Mathematical Olympiad, solving five out of six problems under competition-like conditions (*8*). Models also improved on benchmarks which measure logical and mathematical reasoning ability, including GPQA Diamond, which contains questions about fields such as biology, physics, and chemistry, and on AIME, a competition-level maths test (*23\**, *30\**, *31*). It is particularly important to monitor improvements in mathematical reasoning, as this capability also improves performance in other domains, such as verifying safety-critical software, solving complex scientific problems, and contributing to AI research itself (*22*, *32*).

There is debate over the extent to which recent improvements in AI models reflect genuine reasoning ability, given current limitations in both AI performance and evaluation approaches. For example, one study found that reasoning models cannot solve problems above certain complexity levels, even when given adequate computational resources at inference time. This suggests that these models' success may rely on sophisticated pattern-matching rather than 'true' reasoning (*33\**). This interpretation is reinforced by findings that reasoning models' performance can be sensitive to which test is used, dropping by as much as 65% when benchmark questions are rephrased (*34\**). In addition, transcripts of these models' intermediate steps reveal inefficiencies such as early fixation on wrong answers. Other studies highlight further limitations, showing that even leading models perform much worse than humans in simple spatial reasoning, such as identifying different views of the same object (*35\**), and that they sometimes produce correct answers through flawed logic (*36*, *37*).

Whether these flaws will limit the practical utility of these new models, and when (or whether) new development techniques will address them, are important open questions. Researchers are working to address these limitations through improved training methods and verification systems (among other methods) (*38\**, *39*, *40\**, *41*). For example, some new systems combine general-purpose AI models with specialised mathematical verification programs that can automatically check whether each generated step in the proof is correct (*42*, *43*, *44\**, *45\**).

It is difficult to understand how accurate and useful the evaluations used to assess AI models are. For example, *data contamination* – the inclusion of evaluation questions in training data – can inflate AI models' evaluation scores (*33\**, *46\**, *47*). Most evaluations are conducted only in English, which limits conclusions about AI models' global performance and may overestimate their capabilities in languages other than English (*48\**, *49*). Current benchmarks may also fail to capture the full complexity of real-world reasoning tasks. For example, maths benchmarks focus on problems with clear answers and established solution methods, but in actual mathematical reasoning, the reasoner often has incomplete information and there are multiple valid approaches (*50*, *51*). This means that strong benchmark performance does not guarantee reliable capabilities in practical applications (*52*, *53\**, *54*).

## AI systems are improving at autonomous operation

One year ago, AI agents – general-purpose AI systems that act independently, use tools, and interact with diverse environments to achieve goals – could only complete small-scale tasks in limited demonstrations. Now, some agents can plan and complete multi-step tasks over extended time horizons, albeit with limitations on reliability and largely in controlled environments. In recent studies, researchers have proposed new methods that would allow AI agents to break goals down into sub-tasks, coordinate across multiple other AI agents, and retain memory across longer projects (*55*, *56*, *57\**, *58*, *59*). In real-world scenarios, AI agents are being deployed in limited ways, for example for Web search, software development, or planning trips; however, their efficacy is variable across applications, and better evaluation frameworks are needed to accurately assess agents' performance in the real world (*60*, *61*, *62*).

One way to measure these improvements in AI agents by tracking the complexity of tasks that AI systems can complete autonomously. For example, one benchmark tracks the '50% time horizon' for a set of software engineering and reasoning tasks, meaning the length of task – as measured by how long it would take a human – that AI systems can complete with 50% reliability. Leading AI performance has improved from 18 minutes to over 2 hours over the past year (*52*, *63*). Preliminary analysis suggests that similar exponential trends may apply in other domains. Some data suggests rates of improvement are similar in visual computer use and full self-driving tasks, though AI systems currently perform worse in these domains and the evidence is less robust (*64*).

## AI systems are now commonly used as coding assistants

Coding capabilities have also advanced particularly quickly. Between late 2024 and mid-2025, general-purpose AI systems progressed from simple assistants to more autonomous agents that can use tools, plan, write code, test, and fix bugs across relatively simple software projects under idealised conditions (*65*, *66*). For example, top models now solve over 60% of the problems in the 'SWE-bench Verified', a database of small-to-medium sized real-world software engineering problems (*67*, *68*). The best models completed only 40% of these tasks in late 2024 and almost 0% at the beginning of 2024.

**Figure 2: General-purpose AI system performance on SWE-bench Verified benchmark**

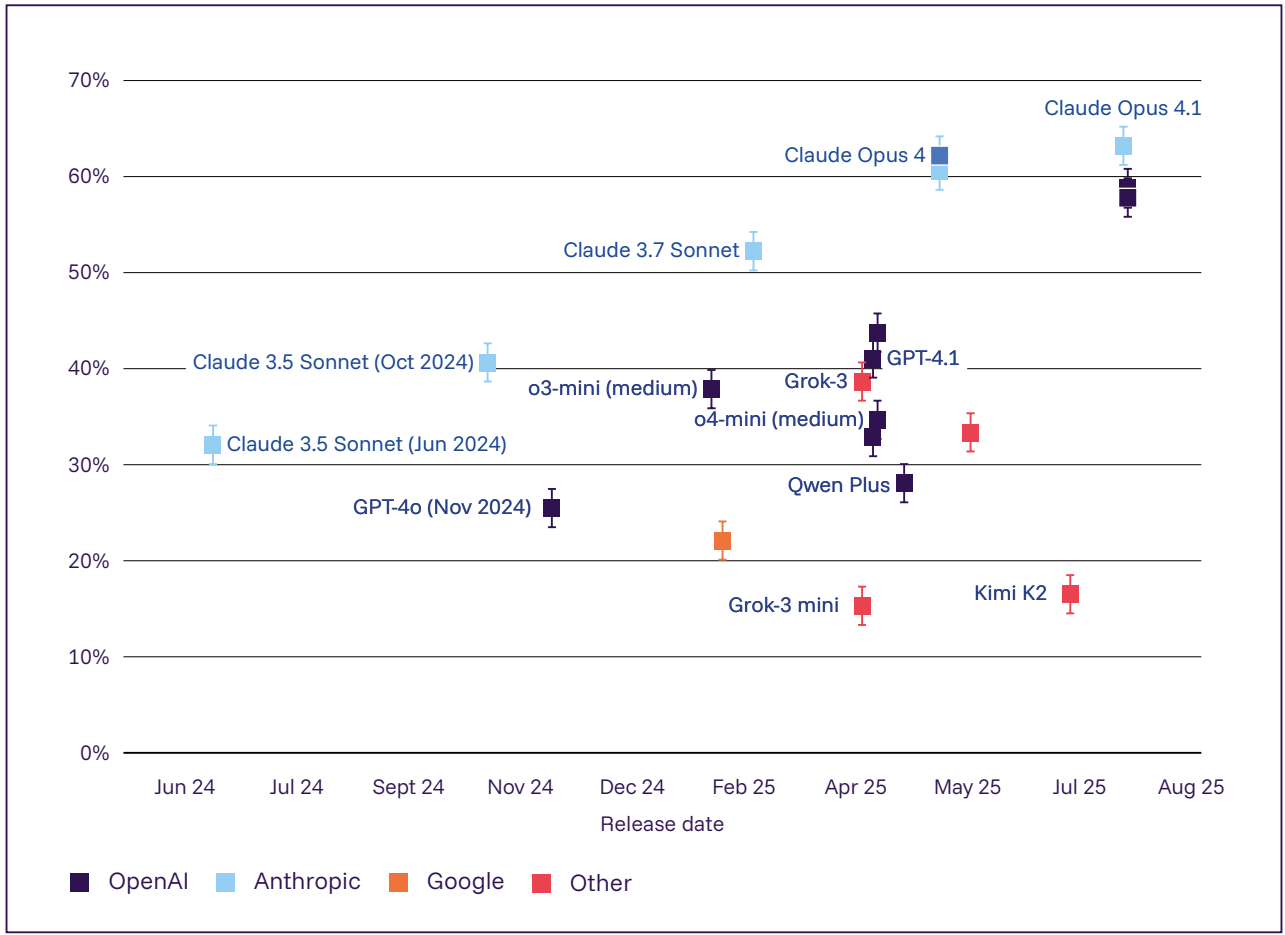

Figure 2: General-purpose AI system performance on 'SWE-bench Verified', a database of real-world software engineering problems. The proportion of problems that the best systems can solve increased from 41% to over 60% in less than a year. Source: Epoch AI, 2025 (*69*).

Benchmark results alone need to be interpreted carefully. Data contamination affects coding benchmarks. A recent analysis of SWE-bench Verified found that models showed up to 35% verbatim text overlap with benchmark problems, indicating that they had memorised benchmark questions during training. In comparison, there was only 18% overlap with similar tasks from other coding benchmarks (*70\**). Similarly, when tested on LiveCode Bench Pro – a benchmark designed to minimise data contamination – the top reasoning model solved 53% of medium-difficulty tasks and 0% of hard tasks when it could not access external tools (*71*). Beyond contamination concerns, code quality issues persist despite task completion improvements.

One study found that AI-generated code runs at least three times slower and uses far more memory than human-written solutions (*72*). Another found that AI code is often more complex, harder to maintain, and less effective on problems requiring deep domain knowledge (*73*). On the whole, benchmarks are limited evaluation settings that do not necessarily reflect the richness of real-world environments.

Adoption of general-purpose AI systems among professional software developers has grown significantly, though trust rates may be low. One recent study estimated that in 2024, 30% of functions† in the programming language Python written by US open source contributors were AI-generated (*74*). A large survey conducted in 2025 found that 51% of professional software developers on Stack Overflow, an online platform, use AI tools daily (*75*). However, trust rates remain low: 47% reported being "somewhat" or "highly" mistrustful of AI tools, and a majority of respondents reported that they do not use more agentic coding systems (*75*).

---

† A function in programming is a self-contained module of code that accomplishes a specific task, such as adding two numbers or counting vowels in a paragraph.

**Figure 3: AI tool use among software developers**

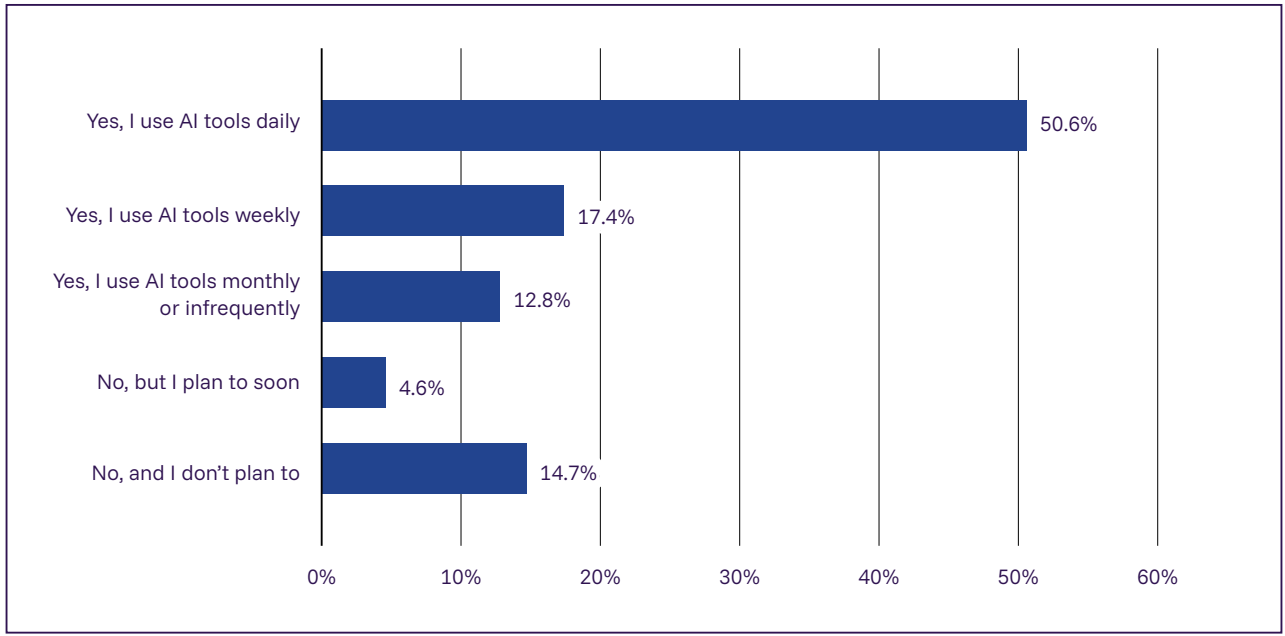

Figure 3: Results from a survey of software developers on AI tool use (n=26,004). A (bare) majority of developers now report using AI tools daily. Source: Stack Overflow, 2025 (*75*).

The effect of AI tools on developer productivity varies significantly across studies and contexts. Large-scale workplace experiments across major companies found that developers with AI code completion tools completed 26% more tasks, with greater benefits for less experienced developers (*76*). However, a smaller controlled study of 16 experienced developers found that, when using AI tools, developers took 19% longer to complete tasks (*77*). This study involved developers working on large, complex codebases they knew well, where their existing familiarity may have made direct implementation faster than coordinating with AI assistance. These varying results likely reflect differences in developer experience, project complexity, and AI tool sophistication. Other studies have found that AI tools can introduce technical debt – coding shortcuts that have immediate benefits but increase long-term maintenance costs – especially when code is integrated without adequate review (*78, 79*). Despite this mixed productivity data, growing adoption and improving capabilities suggest that AI is starting to play a larger role in software development workflows.

# AI systems still underperform on many realistic workplace tasks

Beyond software engineering tasks, performance in actual office environments remains limited. In customer service simulations that domain experts judged realistic in 90% of cases, the best AI agents completed fewer than 40% of tasks (*4*). Similarly, when acting in a simulation of a small software firm, the best agents completed only 30% of 175 workplace tasks such as information gathering and email communication (*80*). These limitations partly reflect the lack of continuity and learning that characterizes effective human collaboration: current AI systems cannot build institutional knowledge or adapt based on ongoing workplace relationships in the way human colleagues do. An evaluation of the ability of general-purpose AI systems to complete open-ended web tasks like planning trips or making purchases found that the best model only succeeded 12% of the time (*5*). Current AI agents exhibit better performance when trained to complement human workers, rather than work autonomously (*6*). A recent study examining the deployment of AI systems highlights that only 5% of task-specific generative AI systems and 40% of general-purpose LLMs are successfully integrated into real-world production (*7*).

# AI systems are more helpful in science

Preliminary evidence shows that scientists are using general-purpose AI systems more, from producing literature reviews to assisting with laboratory work. For example, a study of human-computer interaction research examined 153 scientific papers where the authors reported that they had used general-purpose AI. It found that scientists use AI systems to understand literature, generate research ideas, and analyse data (*11*). While adoption patterns differ across research fields, similar applications are being reported in other scientific domains (*81*). Planning and Web search capabilities together allow AI systems to synthesise findings from diverse sources and produce literature reviews on specific topics (*82\**). There is also more evidence of AI systems assisting in laboratory settings, with general-purpose AI systems helping to design experiments and write protocols in genetics, biomedical, and chemical research (*83*, *84*, *85*, *86*, *87*). An analysis of 15 million biomedical abstracts found that at least 13.5% of publications in 2024 bore stylistic markers of AI use, with the proportion reaching 40% in some disciplines (*88*).

**Figure 4: Frequency of words associated with AI usage in scientific abstracts over time**

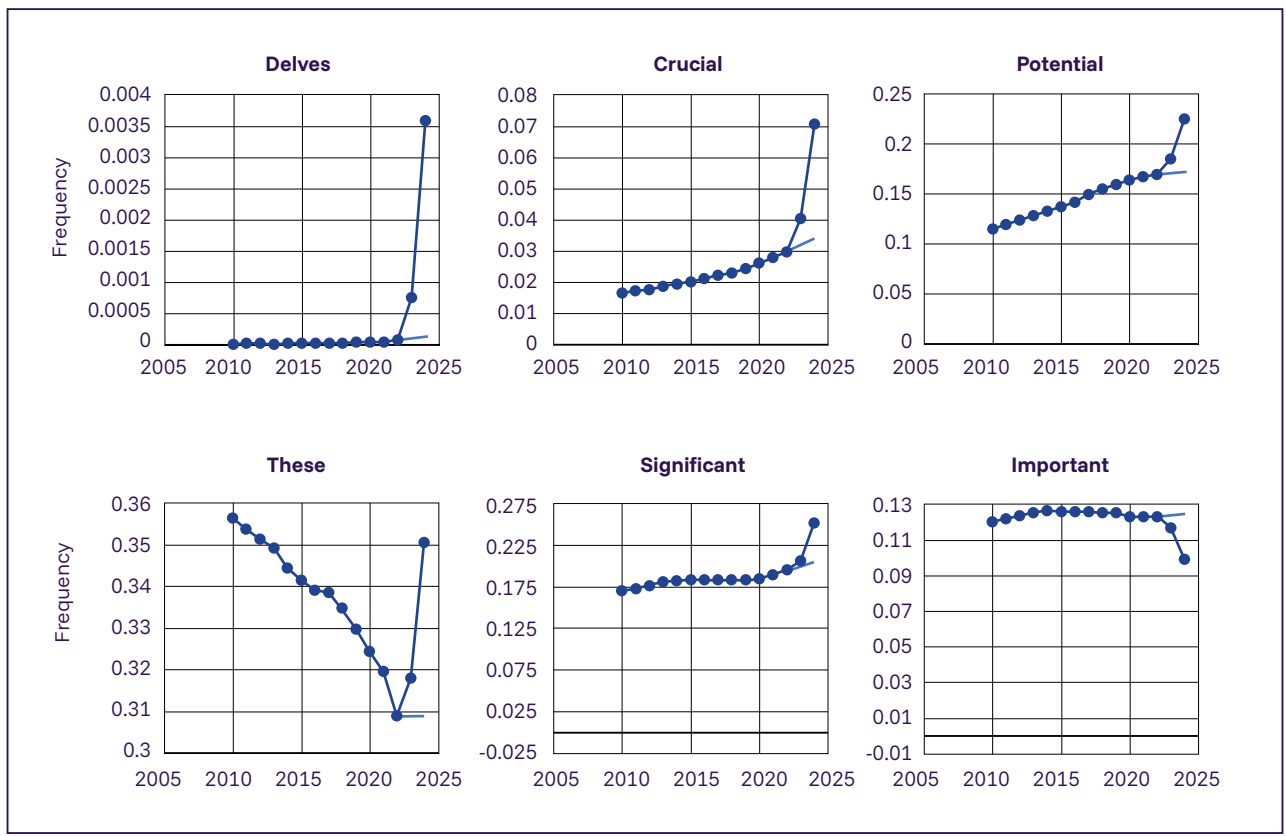

Figure 4: Frequency of words associated with AI usage in scientific abstracts over time. The change in the last two years is evidence of AI assistance in scientific writing and communication. Source: Kobak et al., 2025 (*88*).

General-purpose AI systems clearly remain a complement to, rather than replacement for, human researchers. One evaluation of an autonomous research system found that the papers it produced contained shallow literature reviews, high experiment failure rates, low numbers of citations, and, occasionally, hallucinated results (*89*). Research ideas generated by AI systems score lower on quality than human-generated ideas that end up being published (*90*). Advances in reasoning, search capabilities, and context windows have made AI systems into useful research assistants, but not yet autonomous scientists.

## Multi-modal capabilities have improved

AI models have also continued to improve in image, audio, and video processing abilities. Models can process up to three hours of continuous video or nine and a half hours of audio in a single request (*30\**). In a recent study, Video MMMU, a benchmark that involves answering questions about videos, found that the best model reached about 65% accuracy while human experts averaged 74% (*91*). At the same time, entirely new capabilities are emerging, with interactive video generation models producing outputs that are noticeably higher quality and more difficult to distinguish from real footage (*92*, *93\**). Audio processing has also advanced, with new transcription models like Voxtral lowering costs while maintaining high accuracy (*94\**). These multimodal capabilities allow general-purpose AI systems to operate in more environments and assist with more diverse tasks.

# Implications for risks

## Key information

— **Improved capabilities, including reasoning abilities and autonomous operation, pose new considerations for AI risk management.** Step-by-step problem-solving techniques, extended operational horizons, and improved tool use create new challenges for oversight, particularly when AI systems operate with less human supervision in high-stakes environments.

— **AI capabilities are uplifting both biological and cyber threats while also strengthening defenses.** Leading models assist with various tasks relevant to assisting in the creation of biological weapons. National authorities predict that AI will make cyber crime more accessible and effective in the coming years. A critical research question is whether improved capabilities will benefit attackers or defenders will benefit more.

— **Though many workers have begun to use AI, the labour market impacts of AI systems remain limited.** Evidence points to some workplace adoption and minimal aggregate employment disruption to date, though some targeted impacts on specific demographics have been documented.

— **Some research shows that AI systems may be able to detect when they are in an evaluation setting and alter their behaviour accordingly.** Studies have documented models producing outputs that can mislead evaluators and showing an ability to distinguish between evaluation and deployment contexts. However, evidence comes primarily from laboratory settings, with significant uncertainty about the implications for real-world deployment scenarios and the difficulties it raises for oversight.

As documented in §2. Capabilities, compared to early 2025, AI systems have better problem-solving abilities, extended operational horizons, and are better at using tools. This creates new considerations for AI risk management and oversight.

In response to these new capabilities, some developers have started proactively implementing stronger safeguards as a precautionary measure when releasing AI models. For example, Anthropic released Claude 4 Opus with AI Safety Level 3 (ASL-3) protections due to its improved capabilities in the chemical, biological, radiological, and nuclear (CBRN) domains. Anthropic was unable to determine that 4 Opus had crossed capability thresholds in these domains that would require ASL-3 protections, but neither could it rule out that further testing would uncover such capabilities (*95\**). OpenAI released GPT-5 and ChatGPT Agent with "High capability" safeguards after being unable to rule out that these models could assist novice actors in creating biological weapons, despite lacking definitive evidence of such capabilities. (*12\**, *96\**).[†] Finally, Google DeepMind released its Gemini 2.5 Deep Think model with additional deployment mitigations after determining that the model's technical knowledge of CBRN risks was sufficient to be considered an early warning sign (*99\**).

Another broad development is that more empirical evidence on the nature and severity of various risks is emerging in both experimental settings and real-world deployments. This section provides an overview of new developments since the content in the last Report was finalised in late 2024, focusing on selected risk areas where significant new evidence has emerged.

# Biological risk

Preliminary evaluations indicate that AI systems could soon assist users to develop biological weapons, though the evidence base remains limited and contested. This could include providing instructions for obtaining and constructing pathogens, simplifying technical procedures, and troubleshooting laboratory errors (*12\**, *95\**, *100\**, *101\**, *102\**). While protocols for bioweapons development may already be publicly available online, AI systems can provide more detailed, tailored, or accessible information. For example, one study showed that current language models can troubleshoot virology lab protocols better than 94% of tested subject experts, drawing on knowledge considered rare by virologists (*103*). Such advice could assist both experts and novices, and many current safeguards can be bypassed, such as if the user claims that they need the information for legitimate research (*104*). AI systems can also design custom proteins – the building blocks of many biological weapons – that bind to human targets far more effectively than natural versions and help make viruses resistant to existing treatments (*105\**, *106*). However, a concrete evidence base is still lacking, with many studies lacking peer-review or independent replication. Evaluations also show that general-purpose AI assistance varies across different stages of weapons development (*95\**, *102\**). There is still significant debate about whether current AI systems would substantially assist realistic threat actors (*107*).

Beyond direct scientific assistance, AI systems are also automating parts of the research process, reducing the expertise required for complex biological work. In some cases, AI 'co-scientists' can now independently handle specific research workflows such as hypothesis generation and experimental design that previously required teams of human experts working for weeks or months (*108*, *109\**). For example, AI systems have replicated complex antimicrobial resistance research and quickly validated new medical treatments (*109\**, *110\**).

---

[†] ASL-3 involves increased internal security measures to prevent model theft and deployment restrictions specifically designed to limit misuse for CBRN weapons development (*97\**). OpenAI's "High capability" safeguards similarly involve enhanced security controls and safeguards against misuse before external deployment (*98\**).

**Figure 5: Number of AI-enabled biological tools over time**

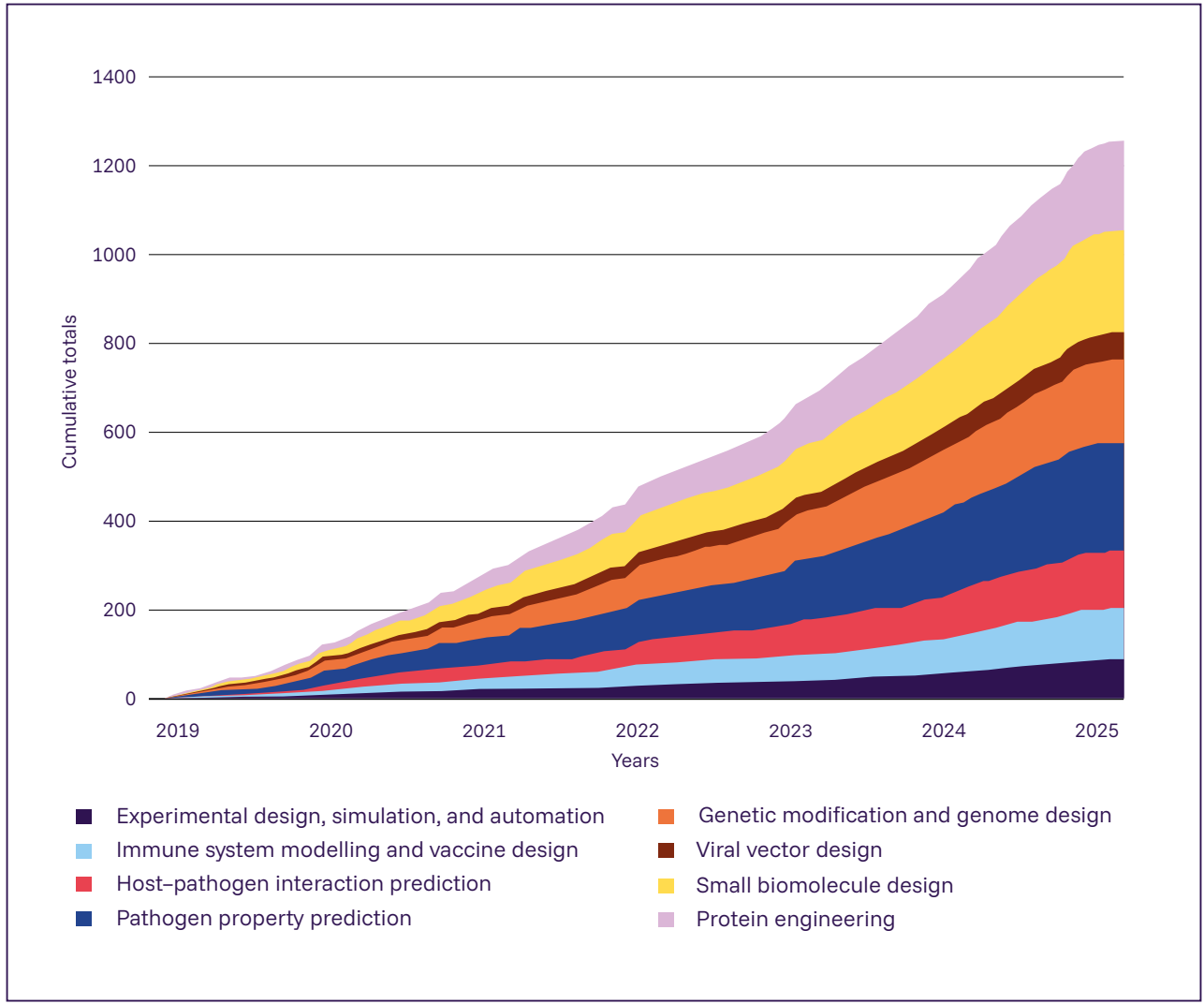

Figure 5: The number of AI biological tools is growing over time. Source: Webster et al., 2025 (*112*).

Cloud laboratories – facilities that allow researchers to conduct automated experiments – are becoming increasingly useful, supported by developments in general-purpose AI. Such laboratories can reduce research timelines from months to hours for some experiments (*111*). While humans remain essential for oversight and implementation, this partial automation and the proliferation of AI tools (see figure 5) mean that some specialised knowledge and laboratory skills that have historically served as barriers to weapons development may become more diffused.

The implications for policy remain uncertain. While laboratory evaluations suggest concerning capabilities, they may not capture the full complexity of actual weapons development environments. As discussed in the Introduction, developers have taken a precautionary approach, implementing additional safeguards on their most capable models despite incomplete evidence about real-world risks (*12\*, 95\*, 96\**).

# Cyber offence and defence

The UK National Cyber Security Centre predicts that by 2027, general-purpose AI systems will almost certainly (95-100% confidence) make cyber offence more effective and efficient, while also offering an opportunity for defence tools (*113*). Consistent with this, evaluations show that AI systems can discover and patch exploitable software flaws and compete with top human teams in hacking competitions (*114*, *115*, *116*, *117*, *118*, *119*). In testing conducted by the Defense Advanced Research Projects Agency (DARPA) AI cyber challenge, one AI system identified 77% of synthetic software vulnerabilities and patched 61% across 54 million lines of code (*118*). As a result, the window to address software vulnerabilities after disclosure has now shrunk to days in some cases, and will likely reduce further as AI advances (*113*). The net effect is that AI could make it cheaper and faster to execute large-scale cyberattacks (*113*, *120*). At the same time, there remain significant weaknesses in AI systems' abilities to independently carry out full attack sequences without human guidance, making human-AI collaboration the primary near-term threat (*113*, *121*, *122*).

In the cyber domain, performance in test environments is translating into real-world impacts, for both beneficial and harmful uses. AI companies report that state-linked and criminal groups are actively using AI models to translate technical sources, analyse disclosed vulnerabilities, develop evasion techniques, and generate code for hacking tools (*123\**, *124\**, *125\**). Europol reports the rise of malicious LLMs on both surface and dark Web, lowering entry barriers for criminal offenders (*126*). These cyber risks may be compounded by the growing use of AI coding assistants across the software development industry, which can introduce security vulnerabilities into widely-used applications (*127*).

At the same time, the ability to identify flaws in code allows cyber-defenders to preemptively patch vulnerabilities before attackers are able to exploit them (*128*, *129*, *130*). It is currently unclear how this 'offence-defence balance' of cybersecurity will evolve given advances in AI capabilities (*131*, *132*). On the one hand, attackers only need to find one critical flaw in order to potentially cause damage, whereas defenders need to be able to find and patch all flaws to guarantee security. On the other hand, attackers commonly need to perform multiple actions in order to complete an attack, each one serving as an opportunity for detection (*131*).

# AI companions

AI companions are increasingly prevalent, and they may pose both risks and benefits to users. Many people are beginning to interact with AI systems more frequently and intimately. Some AI companion applications are AI systems designed to form ongoing personal relationships with users through extended conversations. Some services of this type report having tens of millions of active users (*133*, *134*). The potential risks in these environments remain underexplored, but likely vary by user group, use case, and software design (*135*, *136*).

While AI companions have potential therapeutic applications for reducing loneliness and depression (*137*, *138*), suggested risks including emotional dependence (*139*, *140*, *141*), reinforcing harmful beliefs (*142*, *143*, *144*, *145*), and reported cases of self-harm (*146*, *147*) highlight serious safety concerns. These risks reflect broader challenges around overreliance and inappropriate relationships with AI systems that are already causing documented harms in current deployments (*148\**).

# Labour market risks

New evidence points to some workforce adoption but minimal aggregate labour market effects of general-purpose AI. Several studies have found evidence of notable, but uneven, adoption of general-purpose AI by workers across sectors, usually on a narrow range of tasks ((*149*, *150\**), see also figure 6). Recent studies have also found evidence of increased productivity from AI adoption in the legal sector (*151*), customer service (*152*), and software development (*76*, *77*). Some research suggests targeted labour impacts on specific demographics. For example, one study found that employment for young workers in AI-intensive roles is potentially declining (*153*). Furthermore, studies have documented a decrease in employment in occupations in which AI can automate novice tasks (*154*) or substitute for human skills such as translation (*155*). However, evidence of broader labour market disruption remains limited, with several studies finding no discernible aggregate impact on employment or wages to date (*156*, *157*).

**Figure 6: Prevalence of occupations in US workforce and frequency of relevant Claude conversations**

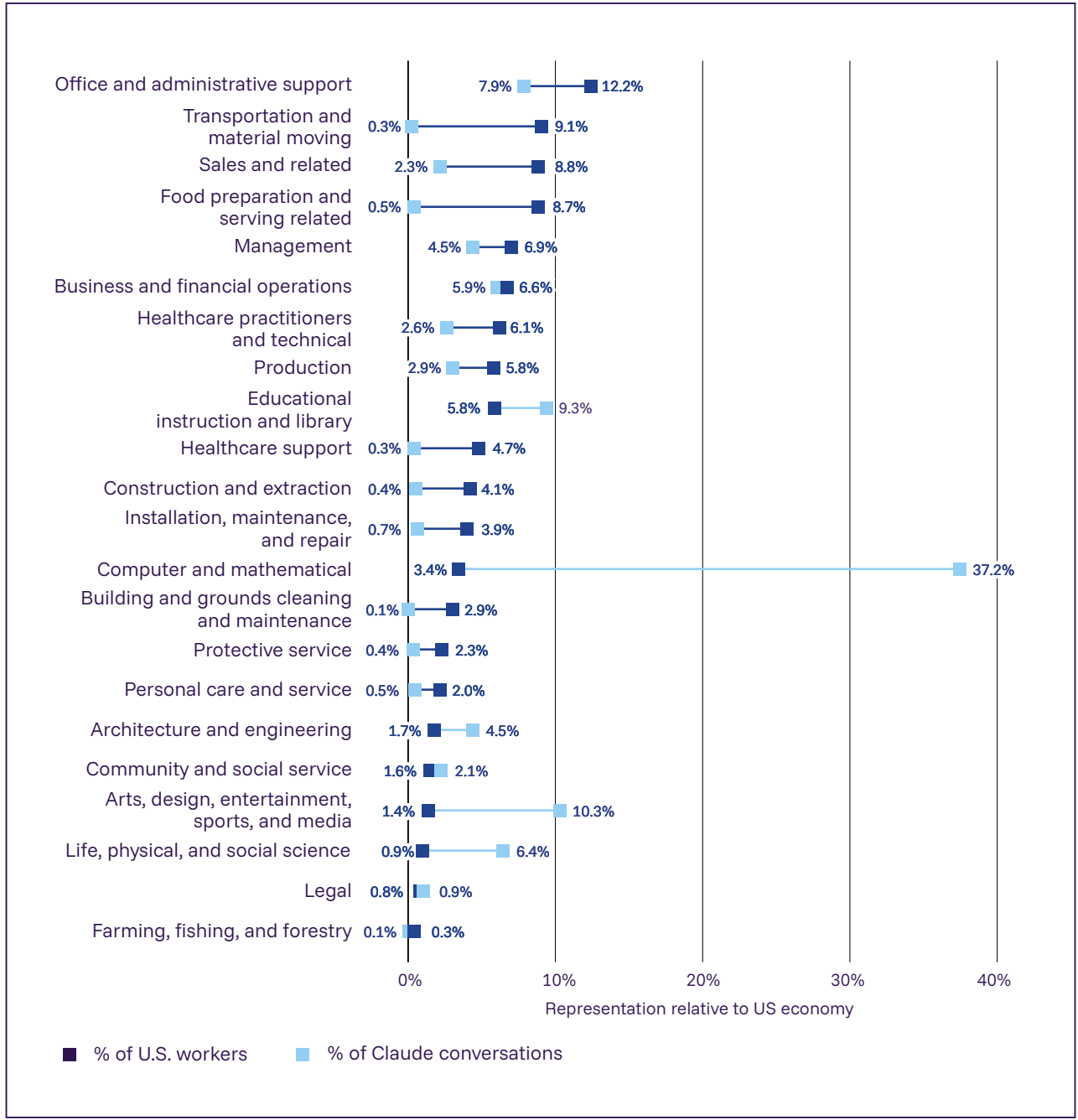

Figure 6: Comparison of how frequently tasks associated with certain occupations appear in user conversations with Anthropic's Claude system, and what percentage of US workers work in those occupations. Usage is highest for professions like software development, and lowest for professions involving physical labour. Source: Handa et al., 2025 (*150\**).

# Monitoring and controllability

Some preliminary research shows that, under certain circumstances, AI systems can detect when they are in an evaluation setting and alter their behaviour accordingly. This creates challenges for monitoring and controlling these systems. Strategic behaviour in evaluation contexts makes it more difficult to predict how AI systems will behave during deployment. This potentially raises the risk of users, AI companies, or other actors losing control of AI systems after deployment. These early research results have prompted researchers to investigate technical measures to assess relevant model propensities and capabilities, and mechanisms for companies to monitor and control their AI systems.

A small number of demonstrations have shown that, under certain conditions, AI models can produce outputs that could systematically mislead evaluators, such as underperforming in assessment contexts (*96\**). This could make it more difficult to assess their true capabilities (*158*, *159\**), though other research finds that these capabilities are not yet sophisticated enough to cause harm during system deployment (*160\**). Since most evidence for these risks still comes primarily from theoretical models and experiments conducted under specific laboratory conditions, there remains significant uncertainty about how likely such behavioural patterns will be in real-world scenarios (*161*).

Work is ongoing into improving the accuracy of evaluations of AI systems. For example, researchers are advancing methods to examine internal components of AI systems in order to better identify concerning behaviours (*162*, *163*). The step-by-step reasoning capabilities of newer models may provide some monitoring opportunities, as their intermediate reasoning steps could potentially reveal concerning behaviours (*164\**). However, the reliability and long-term viability of this oversight approach remains an open research question. For example, recent research has demonstrated that stated reasoning steps do not always accurately represent the model's true reasoning (*165\**, *166\**, *167*). Other researchers are developing alignment techniques aimed at ensuring that AI systems remain responsive to human oversight (*168*).

# Key definitions

— **Capabilities:** The range of tasks that an AI system can perform, and how competently it can perform them.

— **Inference-time enhancements:** Techniques used to improve an AI system's performance after its initial training, without changing the underlying model. This includes clever prompting, sampling multiple responses and choosing the majority answer, using chain of thought, and other forms of *scaffolding*.

— **Inference:** The process in which an AI generates outputs based on a given input, thereby applying the knowledge learnt during training.

— **AI agent:** A general-purpose AI which acts to achieve goals, possibly using plans, adaptively performing tasks involving multiple steps and uncertain outcomes along the way, and interacting with its environment – for example by creating files, taking actions on the web, or delegating tasks to other agents – with little to no human oversight.

— **Evaluations:** Systematic assessments of an AI system's performance, capabilities, vulnerabilities or potential impacts. Evaluations can include benchmarking, red-teaming and audits and can be conducted both before and after model deployment.

— **Benchmark:** A standardised, often quantitative test or metric used to evaluate and compare the performance of AI systems on a fixed set of tasks designed to represent real-world usage or quantify inappropriate behaviour.

— **Control:** The ability to exercise post-training oversight over an AI system and adjust or halt its behaviour if it is acting in unwanted ways.

# References

An asterisk (*) denotes that the reference was either published by an AI company or at least 50% of the authors of a preprint have a for-profit AI company as their affiliation.

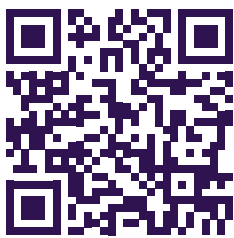

designbysoapbox.com